# On melting of silicon carbide under pressure

Petr S. Sokolov, Vladimir A. Mukhanov, Thierry Chauveau and Vladimir L. Solozhenko*

*LSPM–CNRS, Université Paris Nord, 93430 Villetaneuse, France*

The melting of silicon carbide has been studied at pressures 5–8 GPa and temperatures up to 3300 K. It has been found that SiC melts congruently, and its melting curve has negative slope of −44±4 K/GPa.

Silicon carbide (SiC) is the only compound in the silicon–carbon binary system [1]. At pressures close to ambient, the temperature of SiC incongruent melting may be evaluated at 3100±40 K [1,2], while data on its melting under high pressure are very limited [3–7] and extremely contradictory (Fig. 1), which does not allow one to make any conclusions about the melting behavior (congruent or incongruent) and the slope of the melting curve. In the present work the pressure dependence of the melting temperature of silicon carbide single crystals has been studied in the 5–8 GPa range both in quenching experiments and *in situ*.

The experiments were performed in a specially designed high-temperature (up to 3500 K) cell [8] of a toroid-type high-pressure apparatus. The details of the cell calibration against temperature and pressure are described earlier [8]. In addition to calibrations by the melting temperatures of Si, NaCl, CsCl, Pt, Rh, $Al_2O_3$ and Ni–Mn–C ternary eutectic, the cell was calibrated by the melting temperature of molybdenum (2950(100) K at 7.7 GPa [9]).

Large (2–3 mm) and small (150–300 μm) single crystals of α-SiC (polytype 2*H*, *a* = 3.081 Å, *c* = 5.039 Å) and β-SiC (polytype 3*C*, *a* = 4.357 Å) produced by the interaction of silicon oxide with carbon black at 2800 K in an argon atmosphere with subsequent annealing in air at 1300 K were used in the experiments.

The SiC melting in 5–8 GPa pressure range was studied by quenching, the isothermal holding time at a desired pressure was 60–90 s, the cooling rate at the initial stage after switching-off the power was ~300 K/s. No signs of chemical interaction between SiC and boron nitride capsule were observed over the whole pressure – temperature range under study. Melting of large single crystals was indicated by a change of the shape and microstructure. In the case of small crystals the appearance of a liquid phase in the system was fixed either by the formation of a strong monolith sample with a pronounced laminar structure after quenching (samples, which did not achieve the melting temperature, remained brittle compacts of relatively large crystals) or *in situ* from a jump of the cell electrical resistance caused by a change of conductivity type from the semiconducting (solid SiC) to the metal one (melt) [3,4,10]. In the second case boron nitride capsules were not used and the sample was in a direct contact with a graphite heater.

The results obtained are shown in Fig. 2. Within the experimental error no differences in melting temperatures between α-SiC and β-SiC were observed. The lattice parameters of silicon carbide samples quenched from different pressures and temperatures are very close, and lines of silicon and carbon are absent in the diffraction patterns. All this is leading us to the unambiguous conclusion that SiC melts congruently in the pressure range under study. The SiC melting curve (dashed line obtained by the least-squares method from the results of all experiments) exhibits negative slope (–44±4 K/GPa), which is indicative of the higher density of silicon carbide melt as compared to the solid phase.

The established fact of the decrease of SiC melting temperature with pressure increase in combination with experimentally observed solid-phase amorphization of silicon carbide at ~30 GPa [11] allows us to assume a presence of the phase transition into a denser structure with coordination number 6 (hypothetical γ-SiC) at pressures of 30–40 GPa.

The authors thank Nicolas Fagnon for his help in the preparation of experiments. Financial support from the Agence Nationale de la Recherche (grant ANR-2011-BS08-018-01) is gratefully acknowledged.

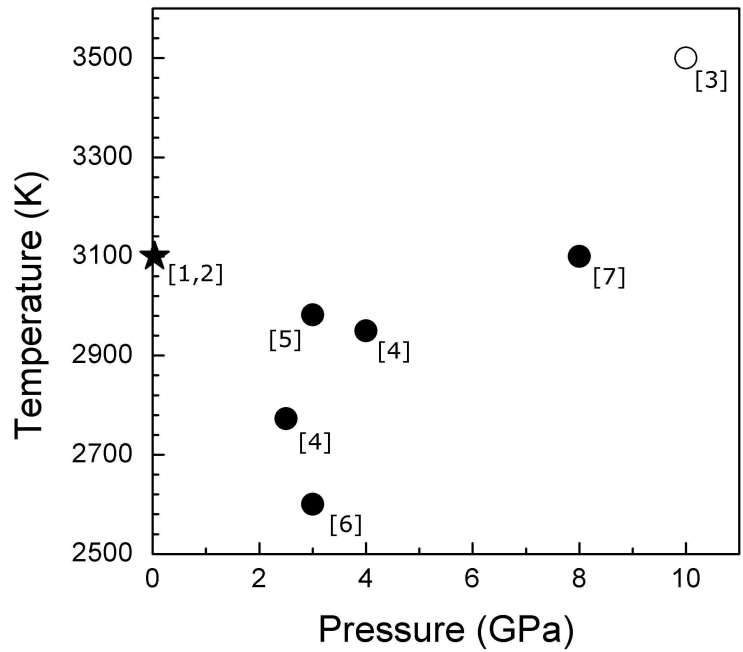

**Fig. 1** SiC melting temperature according to the literature data [1–7]. Solid symbols correspond to the incongruent melting, open symbol indicates the congruent melting. In papers [3–7] the errors of melting temperature determination were not estimated.

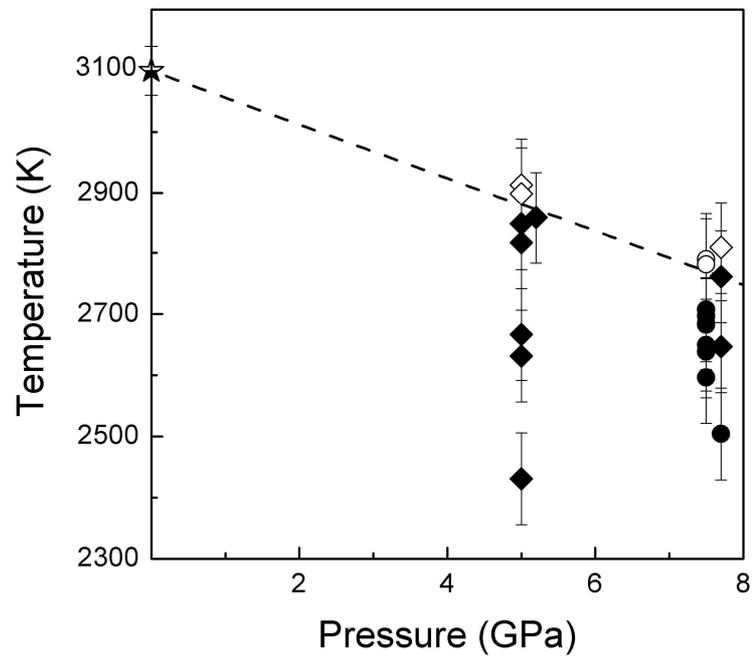

**Fig. 2** Pressure dependence of the SiC melting temperature. The star shows the melting temperature at ambient pressure (3100±40 K [1,2]); diamonds indicate the results of quenching experiments with large single crystals, and circles indicate the results of *in situ* experiments (jump of electrical resistance). Open symbols correspond to the melt, solid – to the solid phase. The dash line is the linear approximation of the melting curve defined by the least-squares method.